\documentclass[proceedings, preprint]{rmaa}

\usepackage{paralist}

\usepackage{psfrag,color}

\SetYear{2008}
\SetConfTitle{Magnetic fields in the Universe II}

\title{Magnetic fields in star formation: from galaxies to stars} 

\author{
  Daniel J. Price,\altaffilmark{1} 
  Matthew R. Bate\altaffilmark{1}
  and Clare L. Dobbs\altaffilmark{1}}

\altaffiltext{1}{University of Exeter, Stocker Rd, Exeter, EX4 4QL (dprice@astro.ex.ac.uk).}

\shortauthor{Price, Bate \& Dobbs}
\shorttitle{Magnetic fields in star formation}

\listofauthors{Daniel J. Price, Matthew R. Bate \& Clare L. Dobbss}
\indexauthor{Price, D. J.}
\indexauthor{Bate, M.R.}
\indexauthor{Dobbs, C.L.}

\abstract{Magnetic fields are important at every scale in the star formation process: from the dynamics of the ISM in galaxies, to the collapse of turbulent molecular clouds to form stars and in the fragmentation of individual star forming cores. The recent development of a robust algorithm for MHD in the Smoothed Particle Hydrodynamics method has enabled us to perform simulations of star formation including magnetic fields at each of these scales. This paper focusses on three questions in particular: What is the effect of magnetic fields on fragmentation in star forming cores? How do magnetic fields affect the collapse of turbulent molecular clouds to form stars? and: What effect do magnetic fields have on the dynamics of the interstellar medium?}

\resumen{ }

\addkeyword{Stars: formation}
\addkeyword{Stars: Pre-main sequence}
\addkeyword{magnetohydrodynamics (MHD)}
\addkeyword{galaxies: magnetic fields}
\addkeyword{galaxies: spiral}

\begin{document}
\maketitle

\section{Introduction}
\label{sec:intro}
 Magnetic fields are observed to be important at every scale in the star formation process--in the dynamics of the ISM in spiral galaxies, during the collapse of individual molecular clouds to form stars, and even down to the fragmentation scale in star forming cores. However we remain significantly ignorant of the role of magnetic fields in constraining the dynamics of collapse at each of these scales.

 In this contribution I will discuss our efforts to incorporate magnetic fields into simulations of the star formation process at these three different scales: in star forming cores, in collapsing turbulent molecular clouds, and in the dynamics of galactic spiral arms. These can be phrased into three unanswered questions regarding the role of magnetic fields in star formation (which we will discuss in order of small$\to$ large scales):
\begin{enumerate}
\item What is the effect of magnetic fields on fragmentation in star forming cores?
\item How do magnetic fields affect the collapse of molecular clouds to form stars?
\item What effect do magnetic fields have on the dynamics of the interstellar medium?
\end{enumerate}

 We have been able to start addressing these three questions in our simulations because of the development over the last 6 years or so (from my [Price's] PhD thesis onwards) of a workable algorithm for incorporating magnetic fields into the Smoothed Particle Hydrodynamics (SPH) method. SPH (for recent reviews see \citealt{price04,monaghan05}) is ideally suited to simulations of star formation because of its Lagrangian nature which means that resolution naturally follows mass. Thus, given that the primary computational challenge of star formation is dealing with the huge range in length and time scales involved during the collapse of a diffuse molecular cloud to form a star, SPH makes a natural choice, though attempts to add magnetic fields have had a long and somewhat tortuous history.
 
\section{Smoothed Particle Magnetohydrodynamics}
  In principle an algorithm for Smoothed Particle Magnetohydrodynamics can be derived straightforwardly from a variational principle \citep{pm04b} starting with the usual hydrodynamic Lagrangian \citep{mp01} with the magnetic energy subtracted:
\begin{equation}
L_{sph} = \sum_{b} m_{b} \left[\frac12 v^{2}_{b} - u_{b} - \frac{1}{2\mu_{0}} \frac{B_{b}^{2}}{\rho_{b}} \right].
\end{equation}
Constraining the Lagrangian using the SPH density summation, the first law of thermodynamics and the SPH representation of the induction equation for the magnetic field leads naturally \citep{pm04b} to the equations of motion for the magnetic fields in SPH in a form which conserves momentum and energy exactly (in practise this means to the accuracy of the time-stepping algorithm). This is of course very beautiful and elegant since it is rare to be able to derive anything, let alone the equations of motion themselves, from a first-principles approach in any numerical method. Unfortunately for life (mine especially) and probably related to some deep conservation law regarding the conservation of beauty (perhaps the divergence-free nature of elegance?), that is where the beauty ends and the fun begins, as there are a number of thorny technical issues which must be dealt with from that point before the algorithm can be deemed `workable'.

 Briefly, these issues are that: i) the exactly-momentum conserving representation of the Lorentz force in SPH turns out to be numerically unstable in the regime where magnetic pressure dominates over gas pressure (the solution is to use an alternative representation which is stable in all regimes but compromises momentum conservation slightly); ii) shocks in MHD are much more complicated than their hydrodynamic counterparts and require careful treatment (the solution is to apply artificial dissipation terms including an artificial viscosity, resistivity and possibly conductivity where appropriate); iii) that the smoothing length in SPH and MHD-SPH is a spatially variable quantity and thus terms involving the derivative of the smoothing length should be accounted for in order to retain the Hamiltonian nature of SPH (ie. exact conservation of energy and entropy). The most thorny of all the technical issues (perhaps the rose bush itself) is that nasty fourth Maxwell equation $\nabla\cdot{\bf B} = 0$ which blights all numerical MHD by the simple fact that it enters the MHD equations as a constraint on the initial condition which should remain satisfied for all time. Whilst divergence cleaning methods similar to those adopted by several grid-based MHD schemes can also be applied in an SPH context (see \citealt{pm05} for details), we have found a more effective approach is to use prevention rather than cure by formulating the magnetic field such that it is divergence-free by construction. A natural choice for a Lagrangian code is to use the `Euler potentials'
\begin{equation}
{\bf B} = \nabla \alpha \times \nabla \beta,
\end{equation}
which is divergence free by construction and has the advantage that the induction equation expressed in terms of the Euler potentials takes the form
\begin{equation}
\frac{d\alpha}{dt} = 0; \frac{d\beta}{dt} = 0;
\end{equation}
corresponding to the advection of magnetic field lines by Lagrangian particles. Note that in two dimensions the Euler potentials formulation is equivalent to a vector potential formulation of the magnetic field where $\alpha \equiv A_{z}$ and $\beta = z$.

Whilst the Euler potentials formulation is a natural and elegant choice, the disadvantage is that in our simulations field growth is lost once there is no longer a clear mapping from the particle distribution at $t=0$ to the present simulation time. For example, imagining two rings of particles carrying the Euler potential representation of a magnetic field, it is obvious that rotating the inner ring through one complete rotation will return the magnetic field to its initial configuration (because the fields have simply been advected), whereas clearly in reality the field should `remember' the winding. Thus field growth can be underestimated when using the Euler potentials formulation.

 The SPMHD algorithm using both the Euler potentials and divergence cleaning methods have been extensively tested against a wide range of problems used to benchmark many grid-based MHD codes in one \citep{pm04a,pm04b,price04,rp07}, two \citep{pm05,rp07} and three \citep{rp07} spatial dimensions.

\begin{figure*}[!t]\centering
  \vspace{0pt}
  \includegraphics[width=0.8\textwidth]{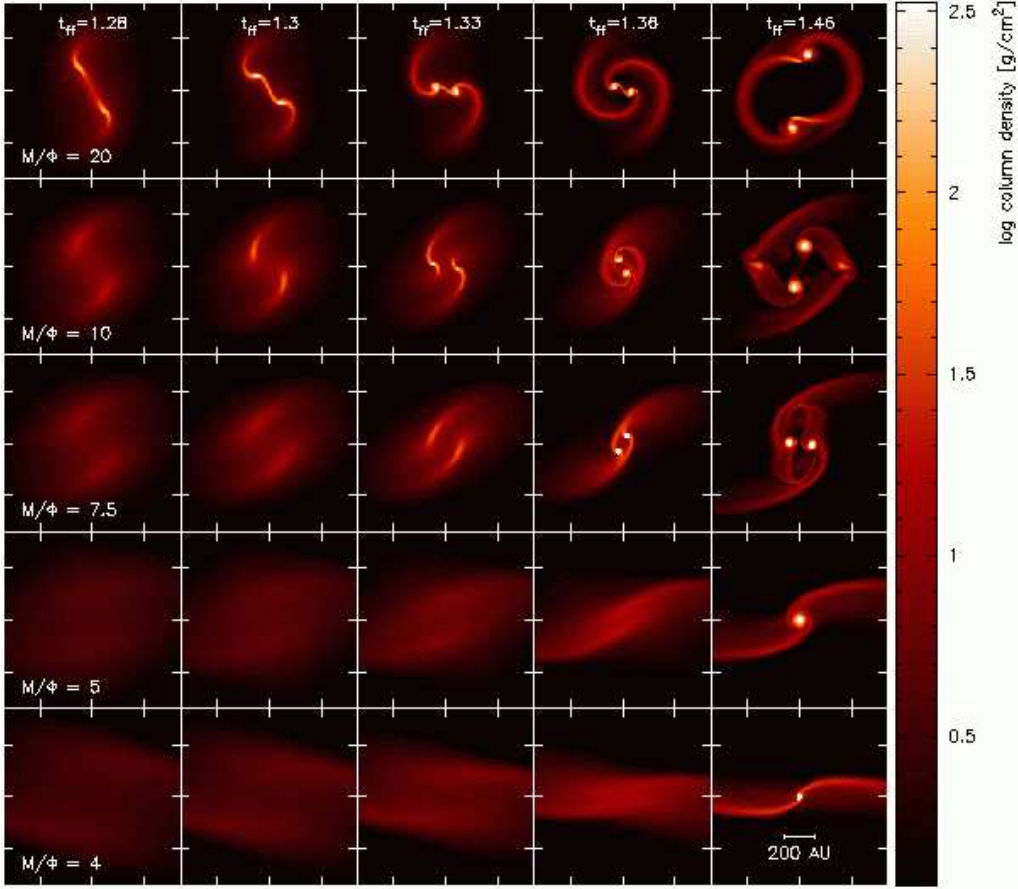}
  \caption{Effect of magnetic fields on fragmentation: simulations of binary star formation using an initial magnetic field perpendicular to the rotation axis.}
  \label{fig:frag}
\end{figure*}

\section{The effect of magnetic fields on fragmentation}
\label{sec:frag}
 One of the first questions we have looked at is: What is the effect of magnetic fields on fragmentation? Whilst some authors have suggested that magnetic fields tend to suppress fragmentation (e.g. \citealt{hw04b}) others have suggested that the presence of magnetic fields may in fact enhance fragmentation (e.g. \citealt{boss02}). To this end we have performed simulations of single and binary star formation in the presence of a magnetic field \citep{pb07} starting from a uniform density $1M\odot$ dense core ($R=4\times 10^{16}$cm)(2674 AU) embedded in a warm, low density medium. The core itself has an initial temperature of 10K, is in solid body rotation and is threaded by an initially uniform magnetic field either perpendicular to or parallel to the rotation axis. We employ a barytropic equation of state which is isothermal until the density reaches 10$^{-14}$g cm$^{-3}$ where it becomes polytropic ($\gamma = 7/5$) in order to mimic the effects of radiative transfer. A binary star formation simulation is initiated by applying a 10\% $m=2$ density perturbation to the otherwise uniform density core.

 The results of a series of calculations with initial magnetic field parallel to the plane of rotation (a $B_{x}$ field in our simulation coordinates) are shown in Figure~\ref{fig:frag}. We find that in general fragmentation is dramatically suppressed in the presence of even relatively weak magnetic fields (mass-to-flux ratios of 10 or lower), leading to single star formation where a binary would have formed in the hydrodynamic case. In these calculations where the initial core rotation is relatively low, the effect is mostly driven by the increase in the effective thermal energy of the cloud (ie. the extra pressure supplied by the magnetic field). This is evident for example if the calculations are re-run using only the magnetic pressure component of the force where a similar suppression of fragmentation is found. Actually the picture is slightly more complicated than that, as may be observed in Figure~\ref{fig:frag}. At $t/t_{ff}=1.36$ (fourth column in Figure~\ref{fig:frag}) the binary separation may be observed to decrease with increasing magnetic field strength until no binary is formed at all. In simulations using a field parallel to the rotation axis (ie. a $B_{z}$ field) this suppression of fragmentation (ie. the formation of only a single star) occurs for even weaker field strengths. However with the field in the plane of rotation as in Figure~\ref{fig:frag} the binary is prevented from merging by a ``magnetic cushion'' which is formed between the two overdense regions by the wrapping up of the magnetic field lines. Thus in this case the tension force actually helps to \emph{promote} fragmentation by preventing overdense regions from merging --- a picture advocated previously by \citet{boss02}. The difference is that although we indeed find that tension can dilute the suppression of fragmentation due to magnetic pressure, the net effect is always that fragmentation is suppressed by the magnetic field relative to hydrodynamics.
 
 We find a similarly dramatic effect on the formation of circumstellar discs in the presence of magnetic fields -- in fact disc formation can be almost completely suppressed even at relatively low magnetic field strengths (mass-to-flux ratios of $\lesssim 5$ -- see \citealt{pb07} for more details). \citet{shuetal06} have suggested that this may simply be a manifestation of the well-known ``magnetic flux problem'' in star formation and therefore that to produce discs one must invoke non-ideal MHD processes (specifically Ohmic resistivity) to dissipate the flux.

\section{The effect of magnetic fields on star cluster formation}
 Given that magnetic fields can have somewhat dramatic effects on the fragmentation of cores, what are the effects on larger scales? In particular how does the presence of a magnetic field affect the decay of turbulence in molecular clouds? How do magnetic fields affect the star formation rate and/or efficiency? What is the effect on the Initial Mass Function from the presence of a magnetic field on larger scales? From a theoretical point of view, how do magnetic fields change the extremely dynamic picture of star formation painted by \citet{bbb03}, where stars compete with each other for accretion from the parent cloud?

\subsection{Key parameters}
 The importance of magnetic fields to this problem can be quantified in terms of three key dimensionless parameters. These are:
\begin{enumerate}
\item The \emph{mass-to-magnetic flux} ratio, expressed in units of the critical value
\begin{equation}
\left(\frac{M}{\Phi}\right) = \frac{M}{\int B {\rm dS}} / \left(\frac{M}{\Phi}\right)_{crit},
\end{equation}
which expresses the competition between magnetic fields and gravity (sub-critical mass-to-flux ratios are stable against collapse, whereas super-critical ratios are unstable).
\item The plasma $\beta$ (ratio of gas-to-magnetic pressure),
\begin{equation}
\beta = \frac{c_{s}^{2} \rho}{B^{2}/2\mu_{0}},
\end{equation}
which expresses the relative importance of magnetic fields as a source of \emph{pressure}.
\item The ratio of RMS turbulent velocities to the Alfv\'en speed
\begin{equation}
f_{turb} = \frac{\sigma_{RMS}}{v_{A}},
\end{equation}
expressing the importance of magnetic fields with respect to turbulence (magnetic fields control turbulence when sub-Alfv\'enic, whereas turbulence controls magnetic fields when super-Alfv\'enic).
\end{enumerate}

 What is not generally appreciated is that these three parameters can be determined independently of each other, since, though they all have a dependance on the magnetic field strength, they depend on this strength to different powers (mass-to-flux ratio depends linearly on field strength, whereas the other parameters depend on the field strength squared) and also on different hydrodynamic parameters (for example changing the sound speed changes only $\beta$ whilst leaving the other parameters unchanged, similarly changing the turbulent RMS velocity amplitude changes only $f_{turb}$). The implication is that magnetic fields can be dominant in one or more senses but not necessarily in the others. For example a cloud can be supercritical (magnetic fields not preventing collapse) and super-Alfv\'enic (magnetic fields not dominating turbulence) but be in the regime where $\beta < 1$ and thus have magnetic fields acting as the dominant source of pressure. Indeed this regime is perhaps the most interesting as observationally it is the one in which molecular clouds appear to lie \citep{crutcher99,ht04}.

\begin{figure*}[!t]\centering
  \vspace{0pt}
  \includegraphics[width=\textwidth]{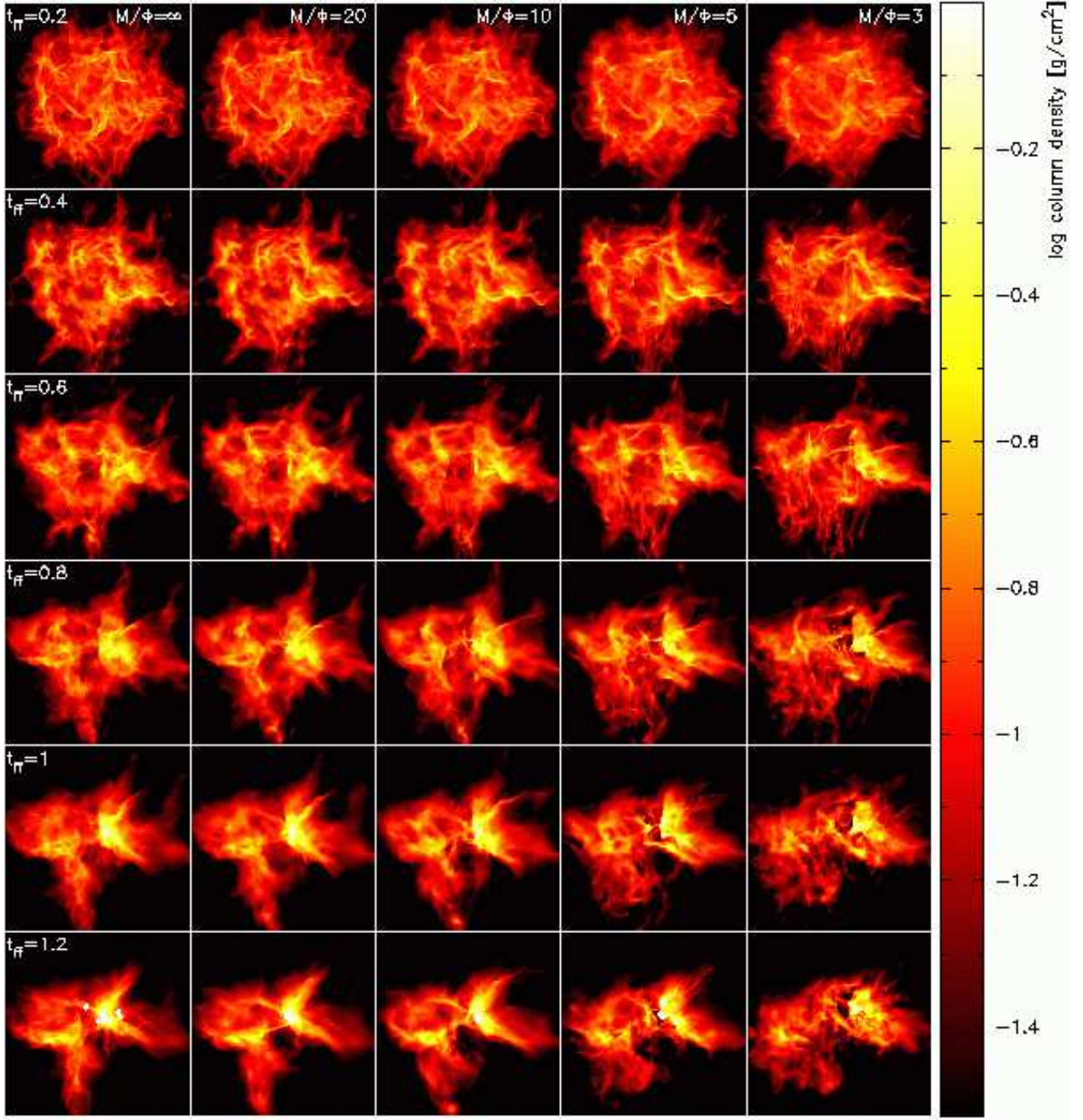}
  \caption{Effect of magnetic fields on star cluster formation: column density in the cloud is shown for the sequence of 5 calculations of progressively increasing magnetic field strength (left to right) at intervals of 0.2 free-fall times (top to bottom). Note in particular the ``stripiness'' in the lower parts of the cloud in the strongest field runs at $t\sim 0.5$ free-fall times and the large-scale voids evident at $t/t_{ff} > 1$.}
  \label{fig:globalcloudev}
\end{figure*}

\subsection{Initial conditions}
 In order to study the effect of magnetic fields on star cluster formation we have performed simulations starting from the same hydrodynamic initial conditions as \citet{bbb03} --- that is an isolated $50 M_{\odot}$ spherical cloud of diameter $0.375$pc (giving $n_{H_{2}} = 3.7 \times 10^{4}$) with a temperature of $10$K and an imposed turbulent velocity field with power spectrum $P(k) \propto k^{-4}$ and RMS Mach number 6.7. The equation of state is again barytropic as in \citealt{bbb03}), which is isothermal down to a density of 10$^{-13}$g cm$^{-3}$ and then becomes polytropic with $\gamma$ then changing with density in order to parameterise the effects of radiative transfer at high densities where the gas becomes optically thick to radiation.

  The magnetic field is imposed as an initially uniform field, with strength parameterised in terms of the mass-to-flux ratio in units of the critical value. We have performed a series of 5 calculations using $M/\Phi = \infty, 20, 10, 5$ and $3$ (the calculations are performed in ideal MHD so runs with subcritical mass-to-flux ratios would not collapse --- rather they bounce and then expand freely). In all cases the turbulent velocities are super-Alv\'enic (by factors ranging from $\infty$ down to $2$ for the 5 runs) meaning that magnetic fields do not control the turbulent velocity field. However of the 5 runs, the two strongest field runs ($M/\Phi = 5$ and $M/\Phi = 3$) have initial plasma $\beta$'s of $0.7$ and $0.25$ respectively, ie. they lie in the regime where magnetic pressure is dominant over gas pressure.

\subsection{Results}
\subsubsection{Cloud structure}
 Results from the 5 simulations are shown in Figure~\ref{fig:globalcloudev}, showing snapshots of column density at intervals of $0.2$ free-fall times (top to bottom) for the runs of progressively increasing field strength (left to right). The initial magnetic field is oriented parallel to the y-axis in this figure.  At first glance the global structure of the cloud is similar in all of the runs, which is a result of the fact that gravity rather than magnetic fields plays the dominant role in determining the initial cloud structure. Some small differences in the development of the initial density fluctuations are visible at $t/t_{ff} = 0.2$ -- namely that the stronger magnetic field runs tend to produce smoother structures due to the additional pressure in the cloud. At later times, however, there are two particularly striking effects which only occur in the two strongest field runs (ie. those with $\beta < 1$).
 
\subsubsection{Magnetically supported voids} 
\begin{figure}[!t]\centering
  \vspace{0pt}
  \includegraphics[width=\columnwidth]{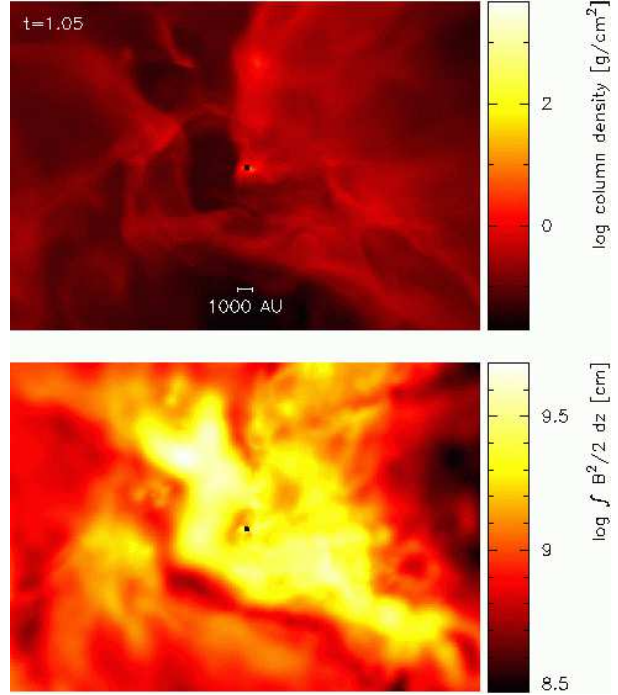}
  \caption{Close-up of the magnetically supported void in the $M/\Phi = 5$ run, showing column density (top) and integrated magnetic pressure (bottom).}
  \label{fig:void}
\end{figure}

 Perhaps the most striking features are the large voids visible in the cloud structure at $t \gtrsim 0.8 t_{ff}$ in the strong $\beta < 1$ magnetic field runs which are completely absent from the hydrodynamic and weak field runs. These voids are directly related to the additional support provided by the magnetic field to the low density material in the cloud, shown in more detail in Figure~\ref{fig:void} which shows a close up of the void structure in the $M/\Phi = 5$ run at $t/t_{ff} = 1.05$. The plot shows the column density in the cloud (top) and the integrated magnetic pressure (bottom) which appears almost as the inverse of the top panel. The physics behind the creation of such ``magnetically-supported voids'' in the simulation is relatively straightforward: since gas is tied to the magnetic field lines, gas motions (due to the presence of a turbulent velocity field) are channelled by the magnetic field to form dense filaments (with relatively higher mass-to-flux ratios and higher $\beta$), leaving behind regions which are relatively lower in density but retain the same field strength. Thus in these regions the gas pressure can be strongly reduced whilst the magnetic pressure remains unchanged, producing an enhancement relative to the gas pressure exactly as observed in Figure~\ref{fig:void}. The creation of the void structure is perhaps better appreciated by viewing an animation of the magnetic field structure evolution overlaid on the column density plots which is a available from the authors' web page\footnote{http://www.astro.ex.ac.uk/people/dprice/pubs/mcluster/}.

Thus the appearance of magnetically supported voids in the cloud structure is an inevitable and generic consequence of turbulence in the presence of magnetic fields which are strong enough to resist compression by the gas motion.

\subsubsection{Column density striations}
 The second effect which the magnetic field imprints on the cloud structure in the $\beta < 1$ regime is visible in the $M/\Phi = 5$ and $M/\Phi = 3$runs at around $t\sim 0.5$ free-fall times as a ``stripy'' structure in the low density outer regions of the cloud (visible in the lower part of the $t/t_{ff} = 0.4$ and $t/t_{ff} = 0.6$ panels in Figure~\ref{fig:globalcloudev}). These ``column density striations'' are a result of streaming motions of the gas along the magnetic field lines. This is nothing more than the well-known anisotropy of turbulent motions in the presence of a strong magnetic field \citet{gs95,vos03} which only becomes visible once the magnetic fields are strong enough to resist distortion by the gas flow and thus channel gaseous material in the direction of the ordered large scale field.

\subsubsection{Effect on the Initial Mass Function}
 The additional support given to the cloud by the magnetic field also has a dramatic effect on the star formation rate, even for relatively weak magnetic fields. For example by $t = 1.5$ free-fall times around the total mass in stars in the $M/\Phi = 5$ and $M/\Phi = 3$ runs is $\sim 4M_{\odot}$ and $\sim 2M_{\odot}$ respectively compared to $\sim 8M_{\odot}$ in the hydrodynamic run. Though we do not follow the evolution of the magnetic field deep into the cores at the resolution of the global simulations due to a somewhat high numerical resistivity (and thus may underestimate the effect of the magnetic field on fragmentation), it is nonetheless interesting to examine what effect the magnetic field support given to the cloud on larger scales has on the mass distribution of stars which form. Given that the cloud only contains 50$M_{\odot}$ of material in the first place, it is not possible to provide meaningful statistics on the detailed form of the initial mass function -- though an effect can still be discerned by looking at cruder statistics such as the ratio of stars to brown dwarfs. 
 
 To this end the number of stars and brown dwarfs and the ratio between the two is given in Table~\ref{tab:starbdratio} for the 5 calculations. We find that in general the presence of the magnetic field tends to reduce the number of low mass objects formed in the simulations. We attribute this effect to the more quiescent nature of the star formation in the stronger field runs, leading to fewer multiple systems and thus fewer ejections (which in the competitive accretion model create the brown dwarfs by preventing any further accretion onto them). This in much better agreement with observations for the substellar IMF \citep{kroupa01,chabrier03} compared to the previous (and more recent high resolution) hydrodynamic calculations which are now known to produce a statistically significant excess of brown dwarfs with respect to observations (Bate, 2008, private communication).
\begin{table}
\begin{center}
\begin{tabular}{|l|l|l|l||}
\hline
$M/\Phi$ & $N_{BDs}$ & $N_{stars}$ & ratio \\
\hline
$\infty$ & 44 & 14 & 3.14 \\
$20$ & 51 & 18 & 2.83 \\
$10$ & 22 & 11 & 2.0 \\
$5$ & 15 & 14 & 1.07 \\
$3$ & 8 & 7 & 1.14 \\
\hline
\end{tabular}
\caption{Ratio of brown dwarfs to stars formed in each of the simulations.}
\label{tab:starbdratio}
\end{center}
\end{table}

\subsubsection{Comparisons with Taurus}
 We have found that, even in magnetically supercritical clouds, magnetic fields can still provide a dominant role in determining cloud structure in the regime where $\beta < 1$. The question to ask is then: What parameter regime is most realistic for the molecular clouds in our own Galaxy in which star formation is observed to occur? Though the determination of sub-critical vs supercritical and super-Alv\'enic vs sub-Alfv\'enic is somewhat controversial on the cloud scale, measurements of magnetic fields in molecular clouds almost without exception indicate that they lie in the regime where $\beta < 1$ \citep{crutcher99,bourkeetal01,ht04}. If this is indeed the case we should therefore expect to find exactly the kind of effects on the cloud structure due to a magnetic field which dominates over the gas pressure that we find in the simulations.
 
  It is with this in mind that we turn to recent observations of the Taurus molecular cloud by \citet{goldsmithetal08}. Two features highlighted by the authors of the survey stand out: striations in the gas column density observed in the $^{13}$CO maps which are extremely well correlated with the direction of the large scale ordered magnetic field and large scale voids (or ``cavities'') in the cloud structure.
The presence of \emph{both} of the features we find in the simulations in the $\beta < 1$ regime strongly suggests that star formation in Taurus is taking place in a regime where magnetic pressures are dominant over gas pressure. This is in broad agreement with direct Zeeman-splitting measurements of field strengths from dense cores in Taurus \citep{crutcher99} giving lower limits of $\beta \gtrsim 0.06$.

  With regards to magnetically-supported voids, in an earlier conference proceedings the same authors \citet{goldsmithetal05} comment on a ``very interesting feature'' at $4^{h}30^{m} + 25^{\deg}$ which ``appears as a hole'', where it appears that ``some agent has been responsible for dispersing the molecular gas''. We suggest that this is a region where magnetic pressure is enhanced relative to gas pressure and that the field geometry should be reasonably ordered with a strong line-of-sight component. The only maps which give any indication of such a feature are the somewhat controversial observations of ``faraday screens'' towards Taurus by \citet{wr04} (see \citet{wr04b} for the detailed Taurus maps) which indeed shown a tantalising enhancement in emission in this area. However more observations are needed to assess our prediction.

\begin{figure*}[!t]\centering
 \includegraphics[width=0.95\columnwidth]{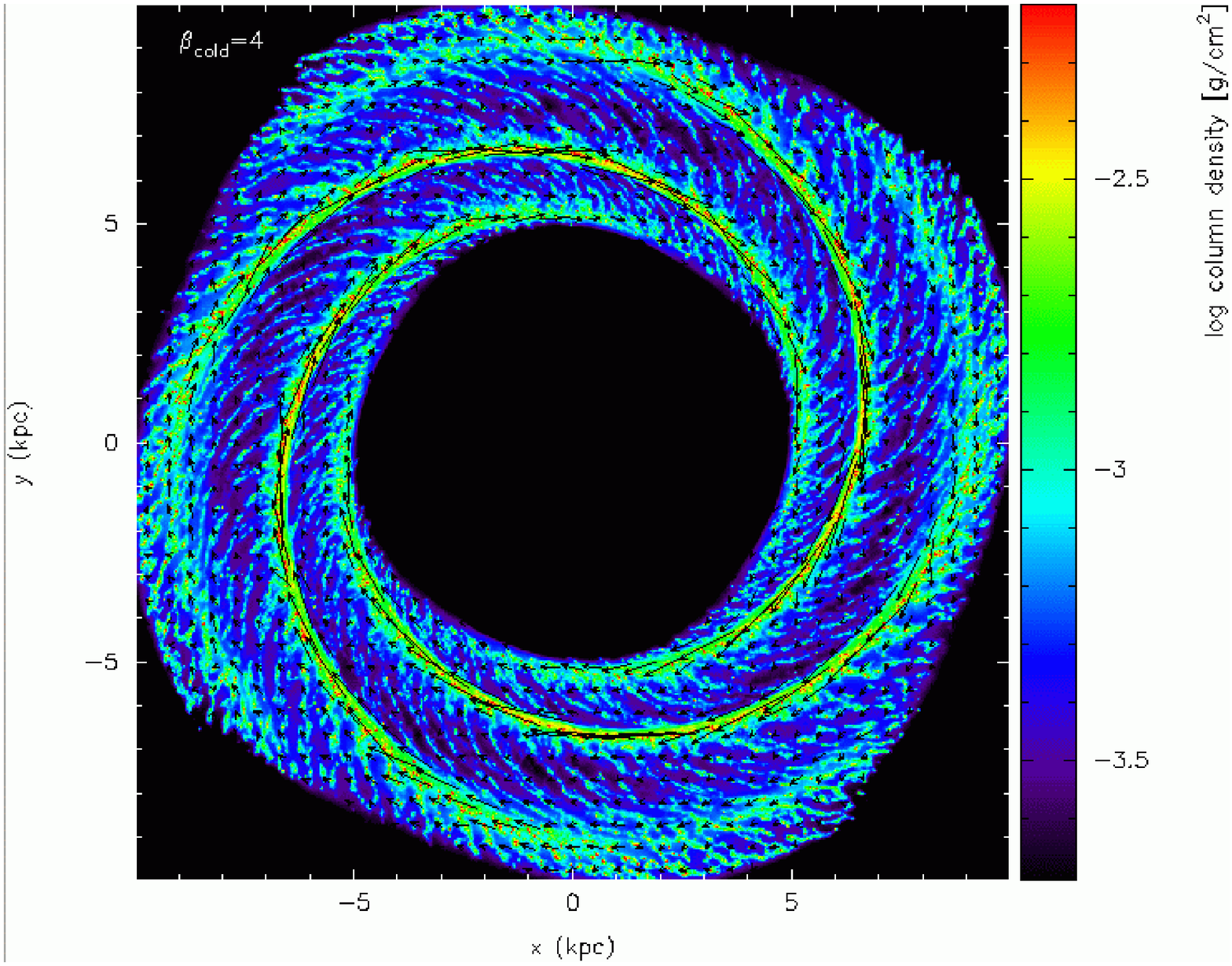}%
 \hspace*{\columnsep}%
  \includegraphics[width=0.95\columnwidth]{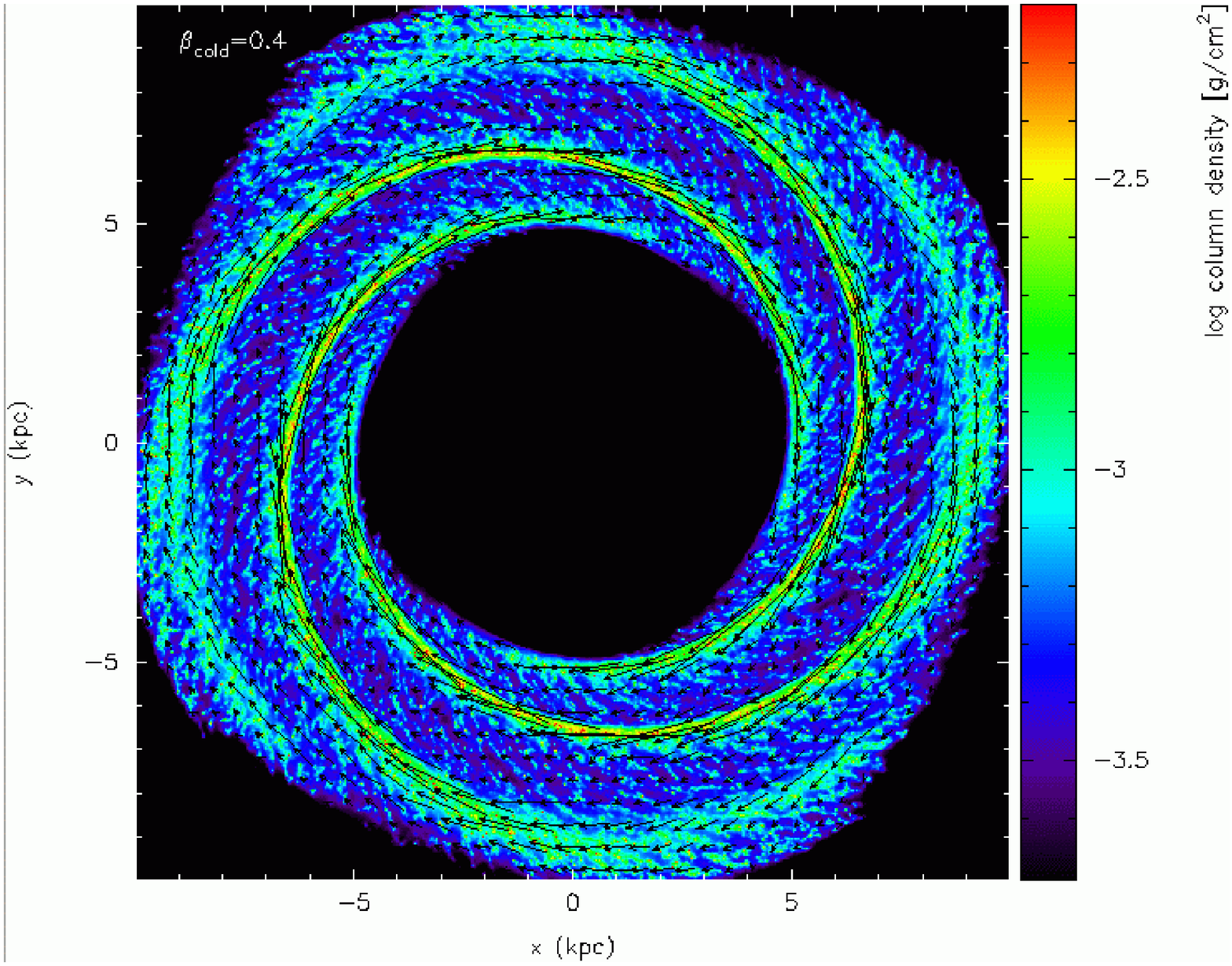}
  \caption{Effect of magnetic fields on the ISM: gas and magnetic field structure in a two-phase ISM simulation using relatively weak (left) and strong (right) magnetic fields. Magnetic fields suppress structure produced by the cold phase of the ISM but do not eliminate it, whilst at weaker strengths the gas is able to induce significantly more disorder in the magnetic field.}
  \label{fig:ism}
\end{figure*}
  
\section{The effect of magnetic fields on the dynamics of the interstellar medium}
 If we are to ever fully understand star formation in molecular clouds then we must make the link to the larger scales in terms of the dynamics of the spiral arms behind which the molecular clouds themselves are formed. On galactic scales magnetic fields also play a significant role in the dynamics of the interstellar medium and it is with this in mind that we have been incorporating magnetic fields into simulations of the ISM on these scales in order to assess both the effect of the magnetic field on the gas flow and vice-versa \citep{dp08}, results of which are shown in Figure~\ref{fig:ism}. We find that the inclusion of a cold phase is crucial in performing simulations of the ISM in galaxies as even relatively small amounts of cold gas produce significant velocity dispersion in the gas and correspondingly induce disorder in an otherwise well-ordered global magnetic field. In general the presence of a magnetic field tends to inhibit the formation of structure in the disc (Figure~\ref{fig:ism}), though provided a cold phase is included in the calculations spurs and spiral arms clumps remain present provided $\beta \gtrsim 0.1$ in the cold gas.
 
\begin{figure}[!t]\centering
  \vspace{0pt}
  \includegraphics[width=\columnwidth]{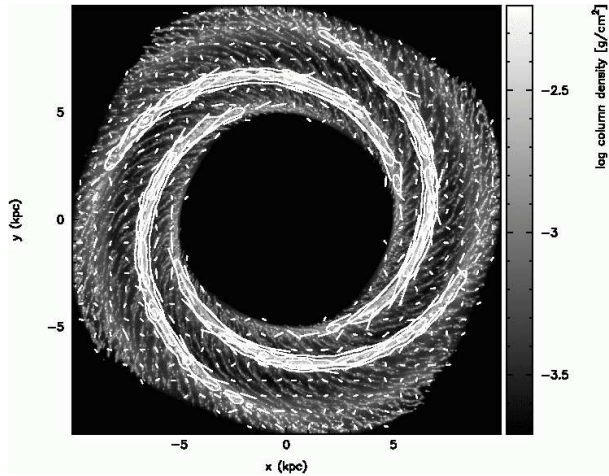}
  \caption{Synthetic polarization map from one of our simulations showing total synchrotron intensity (contours) and the polarisation B-vectors (lines) superposed on a column density map.}
  \label{fig:polmap}
\end{figure}

 The link to observations can be made by producing synthetic polarisation maps based on the magnetic field and gas distributions in our simulations. An example is shown in Figure~\ref{fig:polmap} which shows contours of the total synchrotron intensity and B-vectors for a representative case. Whilst the enhancement in total synchrotron emission in the spiral arm agrees with observations of magnetic fields in spiral galaxies, we also find a corresponding enhancement in the ordered component of the field (though production of ordered fields in the interarm regions is lost when using the Euler potentials formulation) which is \emph{not} observed, suggesting that further work is needed to understand the interaction between spiral arms and magnetic fields in the ISM.

\section{Summary}

\paragraph{What is the effect of magnetic fields on fragmentation?} We have found that the net effect of the magnetic field is always to \emph{suppress} fragmentation, driven by the extra pressure supplied by the magnetic field in a collapsing core. However, magnetic tension can act to dilute this effect to some extent, depending on the field geometry, by preventing overdense regions from merging (an effect we refer to as ``magnetic cushioning''). Magnetic fields also have a dramatic effect on circumstellar disc formation, though we caution that this may be an artifact of employing ideal-MHD calculations in a regime where non-ideal effects are almost certainly important.
\paragraph{How do magnetic fields affect the collapse of molecular clouds to form stars?} Even magnetic fields which do not prevent clouds from collapsing can have a significant effect on the cloud structure in the regime where magnetic pressure dominates over gas pressure (where molecular clouds are ubiquitously observed to lie). In this case we find large scale magnetically-supported voids and striations in column density imprinted on the cloud structure (both of which have been recently observed in the Taurus molecular cloud), a dramatic reduction in the star formation rate and an effect on the IMF in the direction of producing fewer low mass objects.
\paragraph{What effect to magnetic fields have on the dynamics of the interstellar medium?} Magnetic fields tend to smooth out ISM structure, though the presence of a cold gas phase can induce significant disorder in the global field in spiral arm shocks. 

\section*{Acknowledgements}
DJP is supported by a Royal Society University Research Fellowship, though parts of this work were undertaken whilst funded by a UK PPARC research fellowship.
MRB is grateful for the support of a Philip Leverhulme Prize and a EURYI Award. This work, conducted as part of the award ``The format
ion of stars and planets: Radiation hydrodynamical and magnetohydrodynamical simulations"  made under the European Heads
 of Research Councils and European Science Foundation EURYI (European Young Investigator) Awards scheme, was supported b
y funds from the Participating Organisations of EURYI and the EC Sixth Framework Programme. Calculations were performed 
on the United Kingdom Astrophysical Fluids Facility (UKAFF) and Zen, the SGI Altix ICE cluster belonging to the Astrophysics group at the University of Exeter. Visualisations were produced using SPLASH \citep{splashpaper}, a visualisation tool for SPH that is publicly available at http://www.astro.ex.ac.uk/people/dprice/splash.
\bibliography{sph,mhd,starformation}

\end{document}